\documentstyle[12pt]{article}
\renewcommand{\thefootnote}{\fnsymbol{footnote}}

\newcommand{\EQ}{\begin{equation}}
\newcommand{\EN}{\end{equation}}
\newcommand{\bea}{\begin{eqnarray}}
\newcommand{\ena}{\end{eqnarray}}
\newcommand{\vs}[1]{\vspace{#1 mm}}

\newcommand{\uda}{\nearrow \kern-1em \searrow}

\newcommand{\PL}[1]{Phys.\ Lett.\ {\bf #1}}

\newcommand{\PR}[1]{Phys.\ Rev.\ {\bf #1}}

\newcommand{\PTP}[1]{Prog.\ Theor.\ Phys.\ {\bf #1}}

\newcommand{\AJ}[1]{Astorophys. \ J.\ {\bf #1}}

\begin{document}
\setlength{\baselineskip}{7mm}

\begin{titlepage}
\setcounter{page}{0}
\begin{flushright}
EPHOU 99-012\\
OU-HET-333   \\
November 1999
\end{flushright}
\vs{6}
\begin{center}
{\Large Absorption rate of the Kerr-de Sitter black hole and \\ 
the Kerr-Newman-de Sitter black hole}

\vs{5}
{\bf {\sc
Hisao Suzuki\footnote{e-mail address:
hsuzuki@particle.sci.hokudai.ac.jp}}
\\
{\em Department of Physics, \\
Hokkaido
University \\  Sapporo, Hokkaido 060-0810 Japan} \\
{\sc Eiichi Takasugi\footnote{
e-mail address: takasugi@het.phys.sci.osaka-u.ac.jp}
 and  Hiroshi Umetsu\footnote{
e-mail address: umetsu@het.phys.sci.osaka-u.ac.jp}}\\
{\em Department of Physics,\\
Osaka University \\
Toyonaka, Osaka 560-0043 Japan}}\\
\end{center}
\vs{6}

\begin{abstract}
By using an analytic solution of the Teukolsky equation 
in the Kerr-de Sitter and Kerr-Newman-de Sitter geometries, 
an analytic expression of the absorption rate formulae 
for these black holes is calculated.   
\end{abstract}
\end{titlepage}

\renewcommand{\thefootnote}{\arabic{footnote}}
\setcounter{footnote}{0}

%%%%%%%%%%%%%%%%%%%%%%
\section{Introduction}

In a series of papers[1][2], we have developed a method to obtain 
an analytic solution of the Teukolsky equation[3] in Kerr geometries, 
the perturbation equation of massless particles in Kerr 
geometries. 
This method is applied to evaluate the gravitational wave 
from a binary of neutron stars[4]. Recently, we extended 
this method to Kerr-de Sitter and Kerr-Newman-de Sitter geometries 
and showed that an analytic solution is similarly obtained[5]. 
It has been shown that the analytic solution obtained in Ref.5 can be 
analytically continued to cover the entire physical region[6]. 
It should be noted that in Kerr-Newman-de Sitter geometries, 
electromagnetic and gravitational perturbations couple to each 
other and thus these particles are excluded. 

In this paper, we evaluate the absorption rate of the Kerr-de Sitter 
and the Kerr-Newman-de Sitter black holes by using the analytic 
solution. We construct the conserved current by evaluating the 
Wronskian, and we obtain an expression of the absorption 
rate. From this, we show explicitly that  super-radiance 
occurs for the boson case similarly to the Kerr geometry case[2].

The paper is organized as follows. In Sec.2, we summarize 
construction of analytic solutions in order to define parameters 
involved in them. 
In Sec.3, we choose the solution which satisfies the incoming 
boundary conditions at the outer horizon of the black hole 
and examine the asymptotic behavior at the de Sitter 
horizon. Then, we derive an analytic expressions of 
the incident, the reflection and the transmission amplitudes. 
In Sec.4, we derive the conserved current from which we 
derive the absorption rate.  A summary is given in Sec.5.

%%%%%%%%%%%%%%%%%%%%%%%

\section{The analytic solution}
\setcounter{equation}{0}

In the Boyer-Lindquist coordinates, the Kerr-Newman-de Sitter metric
has the form
\bea
ds^2 &=& -\rho^2
\left(\frac{dr^2}{\Delta_r}+\frac{d\theta^2}{\Delta_\theta}\right)
-\frac{\Delta_\theta \sin^2\theta}{(1+\alpha)^2 \rho^2}
[adt-(r^2+a^2)d\varphi]^2 \nonumber \\ 
&&  +\frac{\Delta_r}{(1+\alpha)^2 \rho^2}(dt-a\sin^2\theta
d\varphi)^2,
\ena
where $\alpha=\Lambda a^2/3$,  
$\rho^2=\bar{\rho}\bar{\rho}^*$, $\bar{\rho}=r+ia\cos\theta$ and 
\bea
\Delta_r&=&(r^2+a^2)\left(1-\frac{\alpha}{a^2}r^2\right)-2Mr+Q^2
\nonumber\\
&=&-\frac{\alpha}{a^2}(r-r_+)(r-r_-)(r-r'_+)(r-r'_-)\;,
\nonumber\\
\Delta_\theta&=&1+\alpha\cos^2\theta\;.
\ena
Here $\Lambda$ is the cosmological constant, $M$ is 
the mass of the black hole, $aM$ is its angular momentum, and $Q$ is 
its charge.
 
Next, we deal with the Teukolsky equation. We assume that 
the time and the azimuthal angle dependence of the 
field are described by $e^{-i(\omega t-m\phi)}$. Then, the 
radial part of the equation with spin $s$ and charge $e$ is  
given by 
\bea
&&\makebox[-10mm]{}\Bigg\{ \ 
\Delta_r^{-s}\frac{d}{dr}\Delta_r^{s+1}\frac{d}{dr}
+\frac{1}{\Delta_r}\left[ (1+\alpha)^2 \left(K-\frac{eQr}{1+\alpha}
\right)^2
- is(1+\alpha)\left(K-\frac{eQr}{1+\alpha}\right) \frac{d\Delta_r}
{dr} \right] \nonumber \\
&& \makebox[-5mm]{}
+4is(1+\alpha)\omega r -\frac{2\alpha}{a^2}(s+1)(2s+1) r^2
+s(1-\alpha)-2iseQ -\lambda_s \ \Bigg\} R_s = 0, \label{eqn:Rr}
\end{eqnarray}
with $K=\omega(r^2+a^2)-am$.
This equation has five regular singularities at  $r_\pm, r'_\pm$ and 
$\infty$ which are assigned such that  
$r_\pm \rightarrow M\pm\sqrt{M^2-a^2-Q^2} \equiv r^0_\pm$ and 
$\displaystyle{r'_\pm \rightarrow \pm\frac{a}{\sqrt{\alpha}}}$ 
in the limit $\alpha \rightarrow 0$ $(\Lambda \rightarrow 0)$. 
In Ref.5, it is shown that 
\bea
\lambda_s=\lambda_{-s}.
\ena 

Next, we define the variables $x$ and $z$ as
\bea
x &=& 1-z=\frac{(r_- -r'_-)}{(r_- -r_+)}\frac{(r-r_+)}{(r-r'_-)}.
\ena
This transformation maps 
the outer horizon $r_+$, the inner horizon $r_-$, the de Sitter
horizon 
$r'_+$, and $\infty$ to 0, 1, $x_r$, and $x_{\infty}$, respectively:
\bea
 x_r &=& 1-z_r=\frac{(r_- -r'_-)}{(r_- -r_+)}
 \frac{(r'_+-r_+)}{(r'_+-r'_-)},\nonumber \\
 x_{\infty} &=& 1-z_{\infty}
 =\frac{(r_- -r'_-)}{(r_- -r_+)}\;. 
\ena

Now, we give the solution of the Teukolsky equation 
which satisfies the incoming boundary condition 
on the outer horizon. Before doing so, we define 
parameters to specify the solution:
\bea
\sigma_+&=&2a_2-s+1, \;\sigma_-=-2a_1-2a_3+1,\nonumber\\
\gamma&=&-2a_1-s+1, \;\delta=2a_2+s+1, \;\epsilon=-2a_3-s+1\;, 
\nonumber\\
\omega_H&\equiv& \gamma+\delta-1=\sigma_+ +\sigma_- -\epsilon
=-2a_1+2a_2+1\;.
\ena
Here
\bea
a_1&=&i\frac{a^2(1+\alpha)}{\alpha} \frac{ 
\left(\omega(r_+^2 + a^2)-am-\frac{eQr_+}{1+\alpha}\right)}
{ (r'_+ - r_+)(r'_- - r_+)(r_- - r_+)} \;,\nonumber\\
a_2&=&i\frac{a^2(1+\alpha)}{\alpha} \frac{ 
\left(\omega(r_-^2 + a^2)-am-\frac{eQr_-}{1+\alpha}\right)}
{ (r'_+ - r_-)(r'_- - r_-)(r_+ - r_-)}\;,
  \nonumber\\
a_3&=&i\frac{a^2(1+\alpha)}{\alpha} \frac{ 
\left(\omega({r'_+}^2 + a^2)-am-\frac{eQr'_+}
{1+\alpha}\right)}
{ (r_- - r'_+)(r'_- - r'_+)(r_+ - r'_+)} \;, 
\nonumber\\
a_4&=&i\frac{a^2(1+\alpha)}{\alpha} \frac{
\left(\omega(r_-'^2 + a^2)-am-\frac{eQr_-'}{1+\alpha}\right)}
{(r_- - r'_-)(r'_+ - r'_-)(r_+ - r'_-)}, 
\ena
and the relation $a_1+a_2+a_3+a_4=0$ is satisfied.

With these parameters, the solution is expressed as
\bea
R^{\nu}_{{\rm in};s}&=&\tilde A_s R^{\nu}_{{\rm in};\{0,1\};s}
\nonumber\\
&=& \tilde A_s 
\left[K_{\nu}(s) R^{\nu}_{\{z_r,\infty\};s}(z)
+K_{-\nu-1}(s) R^{-\nu-1}_{\{z_r,\infty\};s}(z)\right],
\ena
where
\bea
R^{\nu}_{{\rm in};\{0,1\};s}&=&
(-x)^{-s-a_1}(1-x)^{a_2}
\left(\frac{x-x_r}{1-x_r}\right)^{-s-a_3}
\left(\frac{x-x_\infty}{1-x_\infty}\right)^{2s+1} 
\nonumber\\ && \times 
\sum_{n=-\infty}^{\infty}a^\nu_n(s)
 F\left(-n-\nu+\frac{\omega_H}{2}-\frac{1}{2},
n+\nu+\frac{\omega_H}{2}+\frac{1}{2};\gamma;x\right)\;,
\nonumber\\
R^{\nu}_{\{z_r,\infty\};s}(z)&&\makebox[-7mm]{}
=z^{a_2}(z-1)^{-s-a_1}
\left(1-\frac{z}{z_r}\right)^{-s-a_3}
\left(1-\frac{z}{z_\infty}\right)^{2s+1} \nonumber \\
&& \makebox[-20mm]{}
\times \sum_{n=-\infty}^{\infty}a^\nu_n(s)
\frac{\Gamma(n+\nu +\sigma_+ +\frac{\omega_H}{2}
+\frac23)
\Gamma(n+\nu-\sigma_- +\frac{\omega_H}{2}
+\frac23)} {\Gamma(2a_2+2a_3+1)\Gamma(2n+2\nu+2)}
\left(\frac{z}{z_r}\right)^{n+\nu+a_1-a_2} \nonumber \\
&& \makebox[-20mm]{}
\times F\left(n+\nu+\sigma_+ -\frac{\omega_H}{2}
+\frac12,n+\nu+\sigma_- -\frac{\omega_H}{2}
+\frac12;
2n+2\nu+2;\frac{z}{z_r}\right),
\ena
and the proportionality constant 
$K_\nu$ is determined by comparing the coefficients of 
 $z^{r+\nu-\frac{\omega_H}{2}+\frac{1}{2}}$ as
\bea
 \label{Knu}
K_\nu \makebox[-3mm]{}&=&\makebox[-3mm]{}
\frac{z_r^{r+\nu-\frac{\omega_H}{2}+\frac{1}{2}}
\Gamma(\gamma)\Gamma(\sigma_+ -\epsilon+1)} 
{\Gamma(r+\nu-\frac{\omega_H}{2}+\frac{3}{2}) 
\Gamma(r+\nu+\delta-\frac{\omega_H}{2}+\frac{1}{2})
\Gamma(r+\nu+\sigma_+-\frac{\omega_H}{2}+\frac{1}{2})
\Gamma(r+\nu+\sigma_--\frac{\omega_H}{2}+\frac{1}{2})} 
\nonumber\\
&& \makebox[-10mm]{}\times
\left[\sum_{n=r}^{\infty} a_n^\nu
\frac{(-)^{n-r}\Gamma(n+\nu-\frac{\omega_H}{2}+\frac{3}{2})
\Gamma(n+\nu+\delta-\frac{\omega_H}{2}+\frac{1}{2})
\Gamma(r+n+2\nu+1)}
{\Gamma(n+\nu+\frac{\omega_H}{2}+\frac{1}{2})
\Gamma(n+\nu+\gamma-\frac{\omega_H}{2}+\frac{1}{2})(n-r)!} \right] 
\nonumber\\
&&\makebox[-10mm]{} \times
\left[\sum_{n=-\infty}^{r} a_n^\nu
\frac{\Gamma(n+\nu-\sigma_+ +\frac{\omega_H}{2}+\frac{3}{2})
\Gamma(n+\nu-\sigma_- +\frac{\omega_H}{2}+\frac{3}{2})} 
{\Gamma(n+\nu+\sigma_+ -\frac{\omega_H}{2}+\frac{1}{2}) 
\Gamma(n+\nu+\sigma_- -\frac{\omega_H}{2}+\frac{1}{2})
\Gamma(r+n+2\nu+2)(r-n)!} \right]^{-1},
\nonumber\\
\ena
which should be independent of $r$, an integer value. 

The coefficients are determined by solving the 
three-term recurrence relation
\bea
\alpha^\nu_n a^\nu_{n+1} + \beta^\nu_n a^\nu_n 
+ \gamma^\nu_n a^\nu_{n-1}=0, 
\ena
where
\bea
\alpha^\nu_n &=& -\frac{(n+\nu-\frac{\omega_H}{2}+\frac{3}{2})
(n+\nu-\sigma_+ +\frac{\omega_H}{2}+\frac{3}{2})
(n+\nu-\sigma_- +\frac{\omega_H}{2} +\frac{3}{2})
(n+\ nu+\delta -\frac{\omega_H}{2}+\frac{1}{2})}
{2(n+\nu+1)(2n+2\nu+3)}, \nonumber\\
\beta^\nu_n &=&\frac{(1-\omega_H)(\gamma-\delta)
(\sigma_+ -\sigma_- +\epsilon-1)
(\sigma_+ -\sigma_- -\epsilon+1)}
{32(n+\nu)(n+\nu+1)}
+\left(\frac{1}{2}-x_r\right)(n+\nu)(n+\nu+1) \nonumber\\
&&     +\frac{1}{4}[\epsilon(\gamma-\delta)+\delta(1-\omega_H)
+2\sigma_+ \sigma_-]
+\frac{\omega_H^2-1}{4}x_r +v, \nonumber\\
\gamma^\nu_n &=& -\frac{(n+\nu+\sigma_+-\frac{\omega_H}{2}
-\frac{1}{2})
(n+\nu+\sigma_- -\frac{\omega_H}{2} -\frac{1}{2})
(n+\nu+\gamma -\frac{\omega_H}{2}-\frac{1}{2}) 
(n+\nu+\frac{\omega_H}{2}-\frac{1}{2})} {2(n+\nu)(2n+2\nu-1)}.
\nonumber\\
\ena
with an appropriate initial condition. Here we set   
$a_0^{\nu}=a_0^{-\nu-1}=1$. Then, we find that  
$a_{-n}^{-\nu-1}=a_n^{\nu}$ is satisfied.

The solution is characterized by the characteristic exponent 
$\nu$ (the shifted angular 
momentum) which is determined so that the coefficients 
are convergent as $n\to \pm \infty$. Since the solution is 
expressed by the sum of hypergeometric functions, 
the convergence of the series is examined. From the behaviors of 
coefficients as $n\to \pm \infty$, we find that 
the solution $R^{\nu}_{{\rm in};\{0,1\};s}$ converges for 
$r<r_+'$ and $R^{\nu}_{\{z_r,\infty\};s}$ converges for 
$r>r_+$, where $r_+$ and $r_+'$ ($r_+ \ll r_+'$) 
are the outer horizon and 
the de Sitter horizon, respectively. 
Therefore, the second expression of $R^{\nu}_{{\rm in};s}$ 
is the analytic continuation of the first expression. 
By combining these two expressions, we can obtain a solution that covers 
the entire physical region.

We now summarize the properties of this solution. We showed that  
$\nu(s)=\nu(-s)$ which is crucial for  solutions with the 
spin $s$ and $-s$ to satisfy the Teukolsky-Starobinsky 
identity[7][8]. 
We showed explicitly that our solution  satisfies 
the T-S identity and as a consequence we can fix the 
relative normalization such that 
\bea
\tilde A_s &=& C_s^* \left[
-\frac{a^2}{\alpha(r_+ -r_-)(r'_+ -r_-)(r'_- -r_-)}\right]^{2s}
\nonumber \\
&& \times \frac{\Gamma(-2a_1+s+1)}{\Gamma(-2a_1-s+1)}
\left| \frac{\Gamma(\nu+a_1+a_2-s+1)}
{\Gamma(\nu+a_1+a_2+s+1)} \right|^2\;(s>0)\;,
\ena
provided $\tilde A_{-s}=1\; (s>0)$. Here, $C_s$ is the 
Starobinsky constant[9].

\section{Asymptotic behavior}
\setcounter{equation}{0}

In general, the asymptotic behavior of the solution 
is found by examining the Teukolsky equation to be[10] 
\bea
\label{eqn:asym-beha}
R_s &\longrightarrow& R_s^{({\rm trans})} \Delta_r^{-s} e^{-ikr_*}, 
           \hspace{1cm} (r \rightarrow r_+) \nonumber \\
    &\longrightarrow& R_s^{{\rm (inc)}} \Delta_r^{-s} e^{-ipr_*}
                      + R_s^{{\rm (ref)}} e^{ipr_*},
           \hspace{1cm} (r \rightarrow r'_+)
\ena
where $k$ and $p$ are defined by 
\bea
k &=& \frac{(1+\alpha)\left[\omega(r_+^2+a^2)-am
                -\frac{eQr_+}{1+\alpha}\right]}
           {r_+^2+a^2}, \nonumber \\
p &=& \frac{(1+\alpha)\left[\omega({r'_+}^2+a^2)-am
                -\frac{eQr'_+}{1+\alpha}\right]}
           {{r'_+}^2+a^2}.
\ena
Here $r_*$ is defined by 
$\displaystyle{\frac{dr_*}{dr}=\frac{r^2+a^2}{\Delta_r}}$ 
and become asymptotically 
\bea
r_* &\longrightarrow& \frac{-ia_1}{k} \ln(-x), 
          \hspace{1cm} (r \rightarrow r_+) \nonumber \\
    &\longrightarrow& \frac{-ia_3}{p} \ln\left(1-\frac{z}{z_r}\right). 
          \hspace{1cm} (r \rightarrow r'_+)
\ena

Here, we give the explicit expressions of $R_s^{{\rm (inc)}}$, 
$R_s^{{\rm (ref)}}$ and $R_s^{({\rm trans})}$ by using 
our analytic solution $R^\nu_{{\rm in};s}$.

From the behavior around $x=1$, we find
\bea
R_s^{({\rm trans})} = \tilde A_s \left[
     \frac{\alpha(r_+ -r_-)^2 (r_- -r'_-)^2(r'_+ -r_-)}
          {a^2 (r_+ -r'_-)}\right]^s
    \left(\frac{-x_r}{1-x_r}\right)^{-a_3}
    \left(\frac{-x_\infty}{1-x_\infty}\right)^{-2s+1}
    \sum_{n=-\infty}^{\infty}a_n^\nu(s).\nonumber \\
\ena
Next, we consider the behavior near the cosmological horizon 
$z \to z_r$. 
First, we consider $R^{\nu}_{\{z_r,\infty\};s}$ and find
\bea
R^{\nu}_{\{z_r,\infty \};s} \longrightarrow 
  A^{\nu}_{+;s} \Delta_r^{-s} e^{-ipr_*}+A^{\nu}_{-;s} e^{ipr_*},
\ena
where 
\bea
A^{\nu}_{+;s} &=& \left[\frac{\alpha(r_+ -r_-)^2(r_- -r'_-)^2(r'_+
-r_-)}
                             {a^2(r_+ -r'_-)}\right]^s
    z_r^{s+a_2}(z_r-1)^{-a_1}
      \left(1-\frac{z_r}{z_\infty}\right)^{-2s+1} \nonumber \\
 && \times \frac{\Gamma(2a_3+s)}{\Gamma(2a_2+2a_3+1)}
      \sum_{n=-\infty}^{\infty}a_n^\nu(s), \nonumber\\
A^{\nu}_{-;s} &=& z_r^{a_2}(z_r-1)^{-s-a_1}
       \left(1-\frac{z_r}{z_\infty}\right)^{2s+1}
        \frac{\Gamma(-2a_3-s) }
            {\Gamma(2a_2+2a_3+1)} \nonumber \\
 && \makebox[-10mm]{} \times 
     \sum_{n=-\infty}^{\infty}a_n^\nu(s)
       \frac{\Gamma(n+\nu+a_3-a_4+1)
             \Gamma(n+\nu-a_1-a_2+s+1)}
            {\Gamma(n+\nu-a_3+a_4+1)
             \Gamma(n+\nu+a_1+a_2-s+1)}.
\ena
Since $R^\nu_{{\rm in};s}$ is expressed by the linear combination of 
$R^\nu_{\{z_r,\infty \};s}$ and $R^{-\nu-1}_{\{z_r,\infty \};s}$,
Eq.(2.9), 
we also need $A^{-\nu-1}_{+;s}$ and $A^{-\nu-1}_{-;s}$.
These are given by using the explicit forms of 
$A^{\nu}_{+;s}$ and $A^{\nu}_{-;s}$ as
\bea
A^{-\nu-1}_{+;s} &=& A^{\nu}_{+;s}, \nonumber\\
A^{-\nu-1}_{-;s} &=& \frac{\sin\pi(\nu+a_3-a_4)\sin\pi(\nu-a_1-a_2+s)}
                          {\sin\pi(\nu-a_3+a_4)\sin\pi(\nu+a_1+a_2-s)}
                     A^{\nu}_{-;s}.
\ena
Now, we obtain the asymptotic amplitudes $R_s^{{\rm (inc)}}$ and 
$R_s^{{\rm (ref)}}$ as 
\bea
R_s^{{\rm (inc)}} &=& 
    \tilde A_s \left[K_\nu(s)+K_{-\nu-1}(s)\right]A^\nu_{+;s}, 
    \nonumber\\
R_s^{{\rm (ref)}} &=& 
    \tilde A_s \left[K_\nu(s)
        +\frac{\sin\pi(\nu+a_3-a_4)\sin\pi(\nu-a_1-a_2+s)}             
{\sin\pi(\nu-a_3+a_4)\sin\pi(\nu+a_1+a_2-s)}K_{-\nu-1}(s)
        \right]A^\nu_{-;s}\;.
        \nonumber\\
\ena

Below, we give some  useful relations to study 
the conserved current and absorption rate,
\bea
\frac{A^\nu_{+;s}}{A^\nu_{+;-s}} &=& 
   \left[\frac{\alpha}{a^2}
       (r'_+ -r_-)^2(r_+ -r_-)(r'_+ -r'_-)\right]^{2s}
   \frac{\Gamma(2a_3+s)}{\Gamma(2a_3-s)}
   \frac{\displaystyle{\sum_{n=-\infty}^{\infty}a_n^\nu(s)}}
        {\displaystyle{\sum_{n=-\infty}^{\infty}a_n^\nu(-s)}}, 
        \nonumber\\
\frac{A^\nu_{-;s}}{A^\nu_{-;-s}} &=& 
   \left[\frac{(r_+ -r_-)(r_- -r'_-)}
              {(r'_+ -r_+)(r'_+ -r'_-)}\right]^{2s}
    \frac{\Gamma(-2a_3-s)}{\Gamma(-2a_3+s)}
\left|\frac{\Gamma(\nu+a_1+a_2+s+1)}{\Gamma(\nu+a_1+a_2-s+1)}\right|^2.
\ena
These lead to the following relations for the asymptotic amplitudes,
\bea
 \label{eqn:asymamp-s}
\frac{R^{({\rm inc})}_s}{R^{({\rm inc})}_{-s}} &=& 
  \frac{1}{C_s}\left[\frac{\alpha}{a^2}
                     (r'_+ -r_+)(r'_+ -r_-)(r'_+ -r'_-)\right]^{2s}
   \frac{\Gamma(2a_3+s)}{\Gamma(2a_3-s)},  
   \nonumber\\
\frac{R^{({\rm ref})}_s}{R^{({\rm ref})}_{-s}} &=&
  C^*_s \left[\frac{\alpha}{a^2}
              (r'_+ -r_+)(r'_+ -r_-)(r'_+ -r'_-)\right]^{-2s}
   \frac{\Gamma(-2a_3-s)}{\Gamma(-2a_3+s)}, \\
\frac{R^{({\rm trans})}_s}{R^{({\rm trans})}_{-s}} &=& 
  \frac{1}{C_s}\left[\frac{\alpha}{a^2}
                     (r_+ -r_-)(r_+ -r'_+)(r_+ -r'_-)\right]^{2s}
   \frac{\Gamma(2a_1+s)}{\Gamma(2a_1-s)}, \nonumber 
\ena
where we used Eq.(4.29) in Ref.6 which is the relation between 
the sums of the coefficients with spin weights $s$ and $-s$.

\section{The conserved current and the absorption rate}
\setcounter{equation}{0}

The conserved current[7] is obtained by examining  
the Wronskian between the incoming solution 
$R^\nu_{{\rm in};s}$ and 
the outgoing solution  
$R^\nu_{{\rm out};s}=(\Delta^{-s}R^\nu_{{\rm in};-s})^*$ 
on the outer horizon. We find
\bea
&& \left[\Delta_r^{s+1}\left(R^\nu_{{\rm in};s}
  \frac{d}{dr}(\Delta_r^{-s}R^\nu_{{\rm in};-s})^*
   -(\Delta_r^{-s}R^\nu_{{\rm in};-s})^*
      \frac{d}{dr}R^\nu_{{\rm in};s}\right)\right]_{r=r_+} \nonumber
\\
&& \hspace{2cm}
 = \left[\Delta_r^{s+1}\left(R^\nu_{{\rm in};s}
   \frac{d}{dr}(\Delta_r^{-s}R^\nu_{{\rm in};-s})^*
    -(\Delta_r^{-s}R^\nu_{{\rm in};-s})^*
      \frac{d}{dr}R^\nu_{{\rm in};s}\right)\right]_{r=r'_+}.
\ena

Then, by substituting the asymptotic behavior (\ref{eqn:asym-beha}),
we find 
\bea
R^{({\rm inc})}_s \left(R^{({\rm inc})}_{-s}\right)^* &=& 
  R^{({\rm ref})}_s \left(R^{({\rm ref})}_{-s}\right)^* 
  \nonumber\\
  && -\frac{(r_+ -r_-)(r_+ -r'_-)(s+2a_1)}
         {(r'_+ -r_-)(r'_+ -r'_-)(s+2a_3)}
      R^{({\rm trans})}_s \left(R^{({\rm trans})}_{-s}\right)^*.
\ena
This relation can be rewritten 
by using the relations (\ref{eqn:asymamp-s})
as
\bea
\left|R^{({\rm inc})}_s \right|^2 &=& \frac{1}{|C_s|^2}
  \left[\frac{\alpha}{a^2}(r'_+ -r_+)(r'_+ -r_-)(r'_+
-r'_-)\right]^{4s}
   \left|\frac{\Gamma(2a_3+s)}{\Gamma(2a_3-s)} \right|^2
    \left|R^{({\rm ref})}_s \right|^2  \nonumber\\
 &&+ \delta_s \left|R^{({\rm trans})}_s \right|^2, 
\ena
where 
\bea
\delta_s = \left[-\frac{(r_+ -r_-)(r_+ -r'_-)}
                       {(r'_+ -r_-)(r'_+ -r'_-)}\right]^{-2s+1}
   \frac{\Gamma(-2a_1-s+1)\Gamma(-2a_3+s)}
        {\Gamma(-2a_1+s)\Gamma(-2a_3-s+1)}.
\ena
For $s=0, \frac12, 1, \frac32$ and 2, $\delta_s$ is explicitly given by  
\bea
\delta_0 &=& 
   \frac{\omega(r_+^2+a^2)-am-\frac{eQr_+}{1+\alpha}}
        {\omega({r'_+}^2+a^2)-am-\frac{eQr'_+}{1+\alpha}}, \nonumber \\ 
\delta_{\frac12} &=& 1,  \nonumber\\
\delta_1 &=& \frac1{\delta_0}\;, 
\nonumber \\
\delta_{\frac32} &=& 
   \left[-\frac{(r'_+ -r_-)(r'_+ -r'_-)}
                {(r_+ -r_-)(r_+ -r'_-)}\right]^2
    \frac{\frac14-4a_3^2}{\frac14-4a_1^2},\nonumber \\
\delta_2 &=& 
   \left[-\frac{(r'_+ -r_-)(r'_+ -r'_-)}
                {(r_+ -r_-)(r_+ -r'_-)}\right]^2
     \frac{(1-4a_3^2)}{(1-4a_1^2)}\frac1{\delta_0}\;. \nonumber
\ena
We recall that the above formulae are valid for all massless 
particles with spin 0, 1/2, 1, 3/2  and 2  
in the Kerr-de Sitter geometry, but for the Kerr-Newman-de Sitter 
geometry, the photon and the graviton are exceptions. 

The relation in Eq.(4.3) 
is interpreted as  the energy conservation[7]. That is, 
the energy balances 
among the incident energy going into the black hole, the energy 
reflected by the black hole, and the energy absorbed by the black hole.
It is evident that the $\delta_s$ are positive definite for fermions, 
because $a_1$ and $a_3$ are purely imaginary. 
For bosons, $\delta_s$ can be negative, and  
super-radiance occurs when 
\bea
\displaystyle{\frac{am+\frac{eQr'_+}{1+\alpha}}{{r'_+}^2+a^2}
<\omega<
\frac{am+\frac{eQr_+}{1+\alpha}}{{r_+ }^2+a^2}}
\ena 
in the Kerr-de Sitter (for massless particles with any spin)[9] and 
the Kerr-Newman-de Sitter (excluding the photon and graviton) 
geometry.

We now give a formula for the absorption rate of the black hole 
in the Kerr-de Sitter geometry using our solution:
\bea
\Gamma_s &=& 1-\frac{1}{|C_s|^2}
  \left[\frac{\alpha}{a^2}(r'_+ -r_+)(r'_+ -r_-)(r'_+
-r'_-)\right]^{4s}
   \left|\frac{\Gamma(2a_3+s)}{\Gamma(2a_3-s)} \right|^2
    \left|\frac{R^{({\rm ref})}_s}
               {R^{({\rm inc})}_s} \right|^2 \nonumber \\
 &=& \delta_s 
    \left|\frac{R^{({\rm trans})}_s}
               {R^{({\rm inc})}_s} \right|^2 \nonumber \\
 &=& \delta_s 
   \left| \frac{(r_+ -r_-)(r_+ -r'_-)}{(r'_+ -r_-)(r'_+
-r'_-)}\right|^{2s}
    \left(\frac{r'_+ -r'_-}{r_+ -r'_-}\right)^2
     \left|\frac{\Gamma(2a_2+2a_3+1)}{\Gamma(2a_3+s)}\right|^2
      \left|K_\nu(s)+K_{-\nu-1}(s)\right|^{-2} \nonumber \\
 &=& \frac{1}{\pi^2 z_r^{2\nu+1}}
      \sin\pi(2a_1+s)\sin\pi(2a_3+s) D^\nu_s 
    \left|1-\frac{p_s^\nu}
                        {\pi^2z_r^{2\nu+1}\sin^2 2\pi\nu}
     D^\nu_s \right|^{-2}, \nonumber \\
\ena
where 
\bea
p_s^\nu&=&\sin\pi(\nu+a_1-a_2)\sin\pi(\nu-a_3+a_4)
                         \sin\pi(\nu+a_1+a_2+s) \sin\pi(\nu+a_1+a_2-s)
                         \nonumber\\
D^\nu_s \makebox[-3mm]{}&=&\makebox[-3mm]{} 
   \left|\Gamma(\nu+a_1-a_2+1)\Gamma(\nu-a_3+a_4+1) \right.\nonumber
\\
 && \times  
  \left.\Gamma(\nu+a_1+a_2+s+1)\Gamma(\nu+a_1+a_2-s+1) \right|^2
d^\nu_s,
  \nonumber \\
d^\nu_s \makebox[-3mm]{}&=&\makebox[-3mm]{} 
  \left|\sum_{n=-\infty}^{0}a^\nu_n(s)
    \frac{\Gamma(n+\nu+a_3-a_4+1)\Gamma(n+\nu-a_1-a_2+s+1)}
         {\Gamma(n+\nu-a_3+a_4+1)\Gamma(n+\nu+a_1+a_2-s+1)
          \Gamma(n+2\nu+2)(-n)!}
    \right|^2 \nonumber \\
 \makebox[-3mm]{}&\times&\makebox[-3mm]{}
  \left|\sum_{n=0}^{\infty}a^\nu_n(s)
    \frac{(-)^n \Gamma(n+\nu+a_1-a_2+1)\Gamma(n+\nu+a_1+a_2+s+1)
                \Gamma(n+\nu+1)}
         {\Gamma(n+\nu-a_1+a_2+1)\Gamma(n+\nu-a_1-a_2-s+1) n!}
    \right|^{-2}\;.
    \nonumber\\
\ena
Here we have set $r=0$, which was an arbitrary 
integer, in $K_\nu(s)$ (\ref{Knu}).
This quantity coincides with that of Ref.2 in the Kerr-limit 
($\Lambda \longrightarrow 0$).

\section{Conclusions and discussions}

We derived an analytic expression of the absorption rate. 
In particular, we found analytically that super-radiance 
occurs for bosons when the frequency satisfies the condition 
given in Eq.(4.7). In order to examine the property in detail, 
we must calculate coefficients by solving the three-term 
recurrence relations in Eq.(2.12). In general, we find 
\bea
\lim_{n\to \infty} \frac{a_{n+1}^\nu}{a_{n}^\nu}=
\lim_{n\to -\infty} \frac{a_{n}^\nu}{a_{n+1}^\nu}=e^{-\xi_r}\;,
\ena
where 
\bea
e^{\xi_r}=1-2x_r+\sqrt{(1-2x_r)^2-1}>1. \qquad (\; x_r<0\;)
\ena
Since $x_r$ is negative and very large for very small $\Lambda$,
\bea
 x_r =\frac{(r_- -r'_-)}{(r_- -r_+)}
 \frac{(r'_+-r_+)}{(r'_+-r'_-)}\simeq \frac{r_-'}
 {2(r_+ -r_-)}\;,
\ena
$e^{-\xi_r}$ is very small. Thus, for larger $n$, the convergence 
of series of coefficients is rapid. In practical cases where 
$\Lambda$ is very small, we first expand $\alpha_n^\nu$, 
$\beta_n^\nu$ and $\gamma_n^\nu$ and also coefficients in terms 
of the small quantity $\alpha\equiv \Lambda a^3/3$, and then we 
expand in terms of $\epsilon \equiv 2M\omega$. In this way, we 
can obtain physical quantities as series in powers of 
$\alpha$ and $\epsilon$. 

At present, we do not know the physical meaning of our analysis 
presented here in comparison with the Kerr geometry case. 
However, we hope that our analysis may become important 
especially when we consider the early universe and also 
when we wish to obtain the deeper insight regarding the correspondence 
between quantum gravity in anti-de Sitter space and conformal 
field theory defined on the boundary\cite{Witten}\cite{Maldacena}.

\newpage

%Appendix format setting
%\setcounter{section}{0}
%\renewcommand{\thesection}{\Alph{section}}
%\renewcommand{\theequation}{\thesection .\arabic{equation}}
%\newcommand{\apsc}[1]{\stepcounter{section}\noindent
%\setcounter{equation}{0}{\Large{\bf{Appendix\,\thesection:\,{#1}}}}}

%\apsc{Various properties}

%%%%%%%%%%%%%%%%%%%%%%%%%%%%%%%%%%%%%%%%%%%%%%%%%%%%%%%%%%%%%%%%%%%%%%%%


\newpage
\begin{thebibliography}{99}
\bibitem{Mano}
S. Mano, H. Suzuki and E. Takasugi, \PTP{95}, 1079 (1996).\\
S. Mano, H. Suzuki and E. Takasugi, \PTP{96}, 549 (1996).
\bibitem{MT}
S. Mano and E. Takasugi, \PTP{97}, 213 (1997).
\bibitem{Teukolsky}
S. A. Teukolsky,
\AJ{185}, 635 (1973).
\bibitem{TTS}
T. Tanaka, H. Tagoshi and M. Sasaki, 
\PTP{96}, 1087 (1996).\\
H. Tagoshi, S. Mano and E. Takasugi, \PTP{98}, 829 (1997).
\bibitem{STU}
H. Suzuki, E. Takasugi and H. Umetsu, \PTP{100}, 491 (1998).
\bibitem{STU2}
H. Suzuki, E. Takasugi and H. Umetsu, \PTP{102}, 253 (1999).
\bibitem{TeukolskyPress}
S. A. Teukolsky and W. H. Press,
\AJ{193}, 443 (1974).
\bibitem{ST}
A. A. Starobinsky and S. M. Churilov,
Sov. Phys. -JETP \bf 38\rm, 1 (1973).
\bibitem{UK} 
Starobinsky constants for Kerr-de Sitter geometries are given by\\
U. Khanal,
\PR{D32}, 879 (1985).
\bibitem{C}
S. Chandrasekhar, {\it The Mathematical Theory of Black 
Holes} ( Oxford University Press (1992) ).
\bibitem{Witten}
E.Witten, Adv. Theor. Math. Phys. {\bf 2}, 253 (1998). 
\bibitem{Maldacena}
J. Maldacena and A. Strominger, \PR{D56}, 4975 (1997).\\
D. Birmingham, I. Sachs and S. Sen, \PL{B413}, 281 (1997).
\end{thebibliography}
\end{document}